\documentclass[preprint,aps,nofootinbib]{revtex4}

\usepackage{graphicx}

\usepackage{amsmath}
\usepackage{amsfonts}
\usepackage{amssymb}
\usepackage{color}

\def\be{\begin{equation}}
\def\ee{\end{equation}}
\def\bea{\begin{eqnarray}}
\def\eea{\end{eqnarray}}

\def\slashchar#1{\setbox0=\hbox{$#1$}           
   \dimen0=\wd0                                 
   \setbox1=\hbox{/} \dimen1=\wd1               
   \ifdim\dimen0>\dimen1                        
      \rlap{\hbox to \dimen0{\hfil/\hfil}}      
      #1                                        
   \else                                        
      \rlap{\hbox to \dimen1{\hfil$#1$\hfil}}   
      /                                         
   \fi}

\begin{document}


\vspace*{1cm}

\title{Polarized Charginos (and Tops) in Stop Decays}

\author{\vspace{0.5cm} Ian Low}
\affiliation{\vspace{0.3cm}
\mbox{High Energy Physics Division, Argonne National Laboratory, Argonne, IL 60439}\\
 \mbox{Department of Physics and Astronomy, Northwestern University, Evanston, IL 60208} \\
 \mbox{Kavli Institute for Theoretical Physics, University of California, Santa Barbara, CA 93106}
 \vspace{0.5cm}}


\begin{abstract}
\vspace{1cm}
Current searches for direct production of scalar top quarks, or stops,  in supersymmetry focus on their decays into $bW^+\tilde{\chi}^0$ by way of  $t  \tilde{\chi}^0$ and $b\tilde{\chi}^+$. While the polarization of the top quark depends on the stop mixing, the chargino turns out to be fully polarized  when the bottom Yukawa coupling can be neglected relative the top Yukawa coupling. We compute the energy and angular spectra of the charged lepton in the chargino channel, which could serve as the spin-analyzer of the chargino. In addition, we demonstrate the top polarization could have a significant impact on the selection efficiencies in direct stop samples at the LHC, while the effect from the chargino polarization is less pronounced.  Two observables in the laboratory frame, the opening angle between the charged lepton and the $b$ quark and the energy of the $b$ quark, are also proposed to optimize searches in the chargino channel versus the top channel.
\end{abstract}

\maketitle


\section{Introduction}


Supersymmetry is the most popular framework for addressing the stability of the electroweak scale. Especially after the discovery of a Higgs boson on July 4th, 2012, the precise mechanism for, or the lack thereof, stabilizing the mass of the Higgs boson has become one of the most outstanding theoretical issues.

 So far experimental searches for  any new particles beyond the standard model have returned null results. Direct search limits for various colored supersymmetric particles such as the gluino and the first two generation scalar quarks, or squarks, have reached the TeV regime at the end of the 7/8 TeV run at the LHC. The only exception is the third generation squarks, which tend to decay to third generation quarks such as the tops and the bottoms and, as a result, have much weaker direct search limits in the vicinity of 500 GeV.
 
Third generation squarks, in particular the stops, play a special role in cancelling the quadratic divergences in the Higgs mass from the top quark. Absence of fine-tuning in the Higgs mass requires the stops to be below 1 TeV and, as such, the LHC will be able to either discover the stops or put strong constraints on so-called "Natural Supersymmetry," where stops are below 1 TeV, by the end of the 14 TeV run.  In other words, experimental searches for light stops have become the litmus test for naturalness in supersymmetry.

At the LHC stops are being searched for in direct production samples $pp \to \tilde{t}_1 \tilde{t}_1^* $, which subsequently decay through either the top channel or the chargino channel \cite{ATLAS-CONF-2012-166, CMS-PAS-SUS-12-023,ATLAS-CONF-2013-061,CMS-PAS-SUS-13-011},
\bea
{\rm Top}&:& \tilde{t}_1 \to t \tilde\chi^0 \to (W^+b)  \tilde\chi^0 \nonumber \ , \\
{\rm Chargino}&:& \tilde{t}_1 \to b \tilde\chi^+ \to b (W^+ \tilde\chi^0) \nonumber \ ,
\eea
both of which yield the same final states. However, it turned out that particular choices of top/chargino polarizations were made in current searches. Moreover, ATLAS and CMS made different choices of polarizations which would lead to different acceptance rates of the possible signals \cite{Claudio}. On the other hand, current searches did not optimize between the top and the chargino channels by utilizing the different decay kinematics, using instead the same selection cuts for both.

Although the possibility of polarized tops in stop decays has long been studied in the literature \cite{Kitano:2002ss, Shelton:2008nq, Perelstein:2008zt, Berger:2012an, Kilic:2012kw, Belanger:2012tm,Bai:2013ema,Belanger:2013gha}, the corresponding issue of chargino polarization seems to have received little attention \cite{Claudio}. In this brief note we compute the energy and angular spectra of the charged lepton in the chargino decay channel of the stop. Much like the charged lepton in the top decays could be used as the spin-analyzer of the top spin \cite{Jezabek:1988ja, Czarnecki:1990pe, Jezabek:1994zv}, the charged lepton in the chargino channel could serve as the spin-analyzer of the chargino. In addition to calculating the lepton spectra, we also study the impact of polarizations on several laboratory frame observables, including those employed in selecting signal events in current searches. Furthermore, two simple observables, the opening angle between the charged lepton and the $b$ quark and the energy of the $b$ quark, are proposed to allow for possible separation of the chargino channel from the top channel in stop decays.

This paper is organized as follows. In Section \ref{sect:2} we spell out the relevant parameters and interaction vertices entering into the stop decays, while the lepton spectra and spin-analyzing powers are presented in Section \ref{sect:spin}. Then in Section \ref{sect:lhc} we perform Monte Carlo studies to study the impact of polarizations on kinematics in the laboratory frame, and propose two simple observables to optimize searches in the chargino versus the top channels. In the last section we conclude.


\section{Polarized Charginos and Tops}
\label{sect:2}


The stop mass-squared matrix in the MSSM in the flavor basis $(\tilde{t}_L, \tilde{t}_R)$ is 
given by \cite{Djouadi:2005gj}
\begin{equation}
\label{stopmass}
M^2_{\tilde{t}} = \left( \begin{array}{cc} 
   m^2_{\tilde{t}_L} + m_t^2 + D_L^t & m_t X_t \\
    m_t X_t &  m^2_{\tilde{t}_R} + m_t^2 + D_R^t
                         \end{array} \right),
\end{equation}
where   
\be
D_L^t = \left(\frac12-\frac23 s_w^2\right) m_Z^2 \cos 2\beta , \quad
D_R^t =  \frac23 s_w^2\, m_Z^2\, \cos 2\beta , \quad
X_t = A_t - \frac{\mu}{\tan\beta} \ .
\ee
In the above $s_w$ is the sine of Weinberg angle. The mass eigenstates $\tilde{t}_1$ and $\tilde{t}_2$ are obtained from
\bea
\left( \begin{array}{cc}
  \tilde{t}_1 \\
  \tilde{t}_2 
       \end{array}   \right) 
      & =&
\left( \begin{array}{cc}
   \cos \theta_t & \sin \theta_t \\
   -\sin \theta_t & \cos \theta_t 
       \end{array}   \right)  
\left( \begin{array}{cc}
     \tilde{t}_L \\
   \tilde{t}_R
       \end{array}   \right)\ , \\
       \sin 2\theta_t &=& \frac{2m_t X_t}{m_{\tilde{t}_1}^2-m_{\tilde{t}_2}^2} \ , \\
\cos 2\theta_t& = &\frac{m_{\tilde{t}_L}^2+D_L^t-m_{\tilde{t}_R}^2-D_R^t}{m_{\tilde{t}_1}^2-m_{\tilde{t}_2}^2} \ .
\eea
where $\theta_t$ is the mixing angle between the flavor basis and mass eigenbasis.

We would like to focus on the two channels most relevant for current stop searches \cite{ATLAS-CONF-2012-166, CMS-PAS-SUS-12-023}:
\be
\tilde{t}_1 \to t\tilde{\chi}^0_1 \to (W^+b)\tilde{\chi}^0_1 \ , \qquad \tilde{t}_1 \to b \tilde{\chi}^+_1 \to b(W^+\tilde{\chi}^0_1) \ ,
\ee
which have identical final states. The relevant interactions in the MSSM lagrangian are \cite{Gunion:1984yn}
\bea
\label{eq:tchi}
{\cal L}_{t\tilde{t}\tilde{\chi}^0}&=&  -\bar{t}\left[y_t N_{j4} P_L +\sqrt{2}\left(\frac{g}{2}N_{j2}+\frac{g'}{6}N_{j1}\right)P_R\right] \tilde{\chi}^0_j \, \tilde{t}_L \nonumber \\
&&\qquad + \bar{t}\left[\frac{2\sqrt{2}g'}3 N_{j1} P_L -y_t N_{j4} P_R \right]\tilde{\chi}^0_j \, \tilde{t}_R \\
\label{eq:bchargino}
{\cal L}_{b\tilde{t}\tilde{\chi}^\pm}&=& \left[ - gV_{i1}\, \tilde{t}_L+  y_t V_{i2}\, \tilde{t}_R \right] \left( \bar{b} \, P_R\, \tilde{\chi}_i^{+\, c} \right) + y_b \,U_{i2}^*\,\tilde{t}_L \left(\bar{b}\, P_L\, \tilde{\chi}_i^{+\, c} \right)   \ , \\
   {\cal L}_{W\tilde{\chi}^+ \tilde{\chi}^0} &=&g W_\mu^- \bar{\tilde{\chi}}_i^0 \gamma^\mu \left[\left(\frac{-1}{\sqrt{2}} N_{i4}V_{j2}^*+N_{i2}V_{j1}^*\right) P_L +\left(\frac{1}{\sqrt{2}} N_{i3}^*U_{j2}+N_{i2}^*U_{j1}\right)P_R\right]\tilde{\chi}_j^+ 
   \label{eq:Wchnu}
\eea    
where $y_t = \sqrt{2} m_t/(v \sin\beta)$ and $y_b = \sqrt{2} m_b/(v \cos\beta)$ are the top and bottom Yukawa couplings. The neutralino and chargino mixing matrices $N_{ij}$, $U_{ij}$, and $V_{ij}$ are defined in the Appendix A in Ref.~\cite{Gunion:1984yn}.

The well-known argument for non-vanishing polarization of top quarks in stop decays is evident in Eq.~(\ref{eq:tchi}), as the top-stop-neutralino coupling is in general parity-asymmetric \cite{Kitano:2002ss, Shelton:2008nq, Perelstein:2008zt}. Moreover, since the top quark decays before it hadronizes, its polarization can potentially be measured through the angular distribution of the decay products \cite{Jezabek:1988ja, Czarnecki:1990pe, Jezabek:1994zv}, which, in turn, can be used to constrain the stop and neutralino mixing angles in the event of discovery. An early proposal considered similar ideas to measure the stau and neutralino mixings using polarized $\tau$ leptons coming out of stau decays \cite{Nojiri:1994it}.

Examining the bottom-stop-chargino coupling in Eq.~(\ref{eq:bchargino}), it is clear that charginos from stop decays are also polarized in general. In particular, in the limit the $b$ quark is massless, $y_b = 0$, the chargino is always in the left-handed {\em helicity} eigenstate in the rest frame of the stop, with its spin pointing opposite to the direction of its motion, much like the $b$ quark. In this limit the chargino is fully polarized independent of the stop mixing parameters.  The effect of finite $b$ quark mass, however, turns out to be enhanced by $\tan\beta$ because of the relation 
\be
\label{eq:ybtan}
y_b = \frac{\sqrt{2}\, m_b}{v \cos\beta} =\frac{\sqrt{2}\, m_b}{v \sin \beta}\tan\beta \approx  \frac{\sqrt{2}\, m_b}{v}\tan\beta
\ee
for $\tan\beta \gg 1$. Therefore $y_b$ becomes as large as $y_t$ when $\tan\beta \alt m_t/m_b$ even though $m_b \ll m_t$.
Similar to the top decays, the chargino polarization can in principle be measured through the angular distribution of its decay products. In this regard, since the $W$-chargino-neutralino interaction in Eq.~(\ref{eq:Wchnu}) depends on both the chargino and neutralino mixing matrices, the chargino channel can be used to probe a different set of mixing entires from those probed by the top channel.

It will be convenient to re-write the relevant interactions in terms of effective couplings in mass eigenstates. Since the top decay is fixed by weak interactions, in the top channel there is only one relevant vertex \cite{Kitano:2002ss, Perelstein:2008zt}
\be
\label{eq:tstopeff}
g_{\rm eff}^{(t)}\ \bar{t} \left( \sin \theta_{\rm eff}^{(t)}\, P_L + \cos \theta_{\rm eff}^{(t)}\, P_R \right) \tilde{\chi}_1^0 \, \tilde{t}_1\ ,
\ee
where 
\be
\tan \theta_{\rm eff}^{(t)} = \frac{y_t N_{j4} \cos\theta_t - \frac{2\sqrt{2}}3 g' N_{j1} \sin\theta_t}{\sqrt{2}\left(\frac{g}2 N_{j2} +\frac{g'}6 N_{j1} \right)\cos\theta_t + y_t N_{j4} \sin\theta_t} \ .
\ee
From the above equation we see that, if the  lightest stop is mostly right-handed (left-handed), $\tilde{t}_1 \approx \tilde{t}_R (\tilde{t}_L)$, then $\sin\theta_{\rm eff}^{(t)} \to 0$  limit can be achieved if the neutralino is mostly a wino/bino (Higgsino), while $\sin\theta_{\rm eff}^{(t)} \to 1$  is obtained for a Higgsino (wino/bino). Therefore it is important to recognize that the top from the stop decays can carry either polarization even if the stop is purely chiral, depending on the nature of the neutralino. In the chargino channel the corresponding vertex is 
\be
\label{eq:bchareff}
g_{\rm eff}^{(\chi)}\ \bar{b} \left( \sin \theta_{\rm eff}^{(\chi)}\, P_L + \cos \theta_{\rm eff}^{(\chi)}\, P_R \right) \tilde{\chi}_1^{+\, c} \, \tilde{t}_1\ ,
\ee
where 
\be
\tan \theta_{\rm eff}^{(\chi)} = \frac{y_b U_{12}^* \cos\theta_t}{-g V_{11} \cos\theta_t +y_t V_{12} \sin\theta_t}  \ .
\ee
Here we see that the chargino polarization is controlled by the left-handed component in the stop. For $\tilde{t}_L$ the chargino is purely right-handed. On the other hand, the vertex for the chargino decay is
\be
\label{eq:chiWeff}
g_{\rm eff}^{(W)} \, W_\mu^- \bar{\tilde{\chi}}_1^0 \gamma^\mu \left( \sin\theta_{\rm eff}^{(W)}\,  P_L +\cos\theta_{\rm eff}^{(W)}\, P_R\right)\tilde{\chi}_1^+ \ ,
\ee
where
\be
\label{eq:tanWeff}
\tan\theta_{\rm eff}^{(W)}=\frac{- N_{14}V_{12}^*+\sqrt{2} N_{12}V_{11}^*}{ N_{13}^*U_{12}+\sqrt{2}N_{12}^*U_{11}}\ .
\ee
In this case the effective mixing angle $\theta_{\rm eff}^{(W)}$ is controlled by the relative components of wino and Higgsino in the chargino as well as the nature of the neutralino. In particular, a polarized $W$ boson would occur only if there is cancellation in either the numerator or denominator in Eq.~(\ref{eq:tanWeff}).


\section{Lepton Spectra in Stop Decays}
\label{sect:spin}


The differential spectra computed below in  Sections \ref{sect:sub1} and \ref{sect:sub2} are, for the most part, either known previously or similar to those of a polarized top decays with anomalous $Wtb$ couplings. They are computed here in order to set the context and double check previous results. In particular, we present the matrix elements of the relevant processes which might be useful should the experimental collaborations wish to re-weight their Monte Carlo simulations in stop searches \cite{Juan}.

For a fermion with the spin vector $s^\mu$, the spin-projection operator is defined as
\be
\hat{\cal S} = \frac12 (1+\gamma^5 \slashchar{s} )  
\ee
such that
\bea
\hat{\cal S} \,u_{s'}(p) = \delta_{ss'} u_s(p) \ , \quad &&  \bar{u}_s(p) u_s(p) =\hat{\cal S}(\slashchar{p} + m) \ , \\
\hat{\cal S}\, v_{s'}(p) = \delta_{ss'} v_s(p) \ , \quad && \bar{v}_s(p) v_s(p) =\hat{\cal S}(\slashchar{p} - m)\ .
\eea
We assume the chargino  (top) in the stop decays is produced on-shell and compute the lepton spectra in stop decays in two stages: $\tilde{t}_1$ decays to $\tilde{\chi}^+_0 b\ ( t \tilde{\chi}^0_1)$ followed by the chargino (top) decays.

\subsection{$\tilde{t}_1\to\tilde{\chi}^+_0 b\ ( t \tilde{\chi}^0_1)$}
\label{sect:sub1}

Given Eq.~(\ref{eq:bchareff}), the amplitude for $\tilde{t}_1\to\tilde{\chi}^+_0 b$ is
\be
i{\cal M} =g_{\rm eff}^{(\chi)} \left[\sin\theta^{(\chi)}_{\rm eff} \bar{u}(p_b) P_L\, {\cal C}\bar{u}(p_{\tilde \chi^+})^T +  \cos\theta^{(\chi)}_{\rm eff} \bar{u}(p_b) P_R\,  {\cal C}\bar{u}(p_{\tilde\chi^+})^T \right]\ ,
\ee
where ${\cal C}$ is the charge-conjugation operator. Using the relation ${\cal C}\bar{u}(p)^T = v(p)$, we have
\bea
|{\cal M}|^2 &=&\left(g_{\rm eff}^{(\chi)}\right)^2 {\rm Tr}\left[(\slashchar{p}_{b}+m_b)(\sin\theta^{(\chi)}_{\rm eff} P_L+ \cos\theta^{(\chi)}_{\rm eff} P_R )\right.\nonumber \\
&&\qquad \qquad \times \left. \hat{\cal S}(\slashchar{p}_{\tilde\chi^+} - m_{\tilde\chi^+})(\sin\theta^{(\chi)}_{\rm eff} P_R+ \cos\theta^{(\chi)}_{\rm eff} P_L ) \right] \nonumber \\
&=&\left(g_{\rm eff}^{(\chi)}\right)^2 \, \left(p_b\cdot p_{\tilde\chi^+}-m_b m_{\tilde\chi^+} \sin 2\theta^{(\chi)}_{\rm eff} -  m_{\tilde\chi^+}\cos2\theta^{(\chi)}_{\rm eff} \ p_b\cdot s \right)\ .
\eea
Specializing to the rest frame of the chargino, we can write
\be
|{\cal M}|^2=\left(g_{\rm eff}^{(\chi)}\right)^2 \, m_{\tilde\chi^+} \left(E_b-m_b  \sin 2\theta^{(\chi)}_{\rm eff} +  \cos2\theta^{(\chi)}_{\rm eff}  |\vec{p}_b| |\vec{s}| \cos\theta_b \right)\ ,
\ee
where $\cos\theta_b\equiv \hat{p}_b\cdot \hat{s}$, and the energy and momentum of the $b$ quark in the rest frame of the chargino is
\bea
\label{eq:ebrest}
E_b &=& \frac1{2m_{\tilde\chi^+}}\left(m_{\tilde t_1}^2 - m_{\tilde\chi^+}^2 - m_b^2\right) \ , \\
\label{eq:pbrest}
 |\vec{p}_b| &=&  \frac1{2m_{\tilde\chi^+}} \lambda^{1/2} (m_{\tilde t_1}^2,m_{\tilde\chi^+}^2 ,m_b^2) \ , \\
\lambda(x,y,z)&\equiv& x^2 + y^2 +z^2 -2 xy -2yz -2xz \ .
\label{eq:lambdafun}
\eea
The sign in front of $\cos\theta_b$ can be determined unambiguously by the following physical argument: when $\sin\theta_{\rm eff}^{(\chi)}=0$ the $b$ quark coming out of the stop decay is always in the left-handed helicity eigenstate according to Eq.~(\ref{eq:bchareff}), with the spin pointing opposite to its direction of motion. Conservation of angular momentum then implies the spin of the chargino to be in the same direction as the  $b$ quark momentum. 

The amplitude for $\tilde{t}_1 \to  t \tilde{\chi}^0_1$ is computed from Eq.~(\ref{eq:tstopeff}) to be
\bea
\label{eq:stoptop}
|{\cal M}|^2 &=& \left(g_{\rm eff}^{(t)}\right)^2 \, \left(p_{\chi^0}\cdot p_t + m_t m_{\chi^0} \sin 2\theta^{(t)}_{\rm eff} - m_t  \cos 2\theta^{(t)}_{\rm eff}  \ p_{\chi^0}\cdot s \right) \nonumber \\
&=&  \left(g_{\rm eff}^{(t)}\right)^2 m_t \left(E_{\chi^0} + m_{\chi^0} \sin 2\theta^{(t)}_{\rm eff} + \cos 2\theta^{(t)}_{\rm eff} |\vec{p}_{\chi^0}|  |\vec{s}|\  \cos\theta_{\tilde \chi^0}\right)\ ,
\eea
The first line in Eq.~(\ref{eq:stoptop}) is the Lorentz invariant expression, while the second line is the result in the rest frame of the top quark and agrees with the existing result in Ref.~\cite{Perelstein:2008zt}. The energy and momentum of the neutralino in the top rest frame are similar to  Eqs.~(\ref{eq:ebrest}) and (\ref{eq:pbrest}) with the replacements $m_b\to m_{\tilde\chi^0}$ and $m_{\tilde\chi^+} \to m_t$. The sign in front of $\cos\theta_{\tilde \chi^0} \equiv \hat{p}_{\tilde \chi^0}\cdot \hat{s}$ can again be fixed as follows: in the $m_{\tilde\chi^0} = 0$ and $\sin\theta_{\rm eff}^{(t)} =0$ limit, the neutralino is always in the left-handed helicity eigenstate. The top quark then has its spin pointing in the same the direction as the motion of the neutralino.

\subsection{$\tilde \chi^+ \to W^+ \tilde\chi^0 \to (l^+\nu) \tilde\chi^0$}
\label{sect:sub2}

Using Eq.~(\ref{eq:chiWeff}), the amplitude-squared for the chargino decay can be written as
\be
|{\cal M}|^2 = \frac{\left(g_{\rm eff}^{(W)}\right)^2}{(p_W^2-m_W^2)^2+m_W^2 \Gamma_W^2} \sum_{r,r'=L, R} W_{\mu\rho} T_{rr'}^{\mu\rho} \ ,
\ee
where
\bea
W^{\mu\rho} &=& P_{\mu\nu}P_{\rho\sigma} {\rm Tr}\left[ \gamma^\nu \slashchar{p}_l \gamma^\sigma P_L \slashchar{p}_{\nu}\right]  \ ,\\
P_{\mu\nu} &=& -g^{\mu\nu} + \frac{p_W^\mu p_W^\nu}{m_W^2} \ , \\
T_{rr'}^{\mu\rho} &=&c_r c_{r'} {\rm Tr}\left[ \gamma^\mu P_r\, \hat{\cal S} (\slashchar{p}_{\tilde\chi^+}+m_{\tilde\chi^+})\gamma^\rho P_{r'}\, (\slashchar{p}_b+m_b)\right] \ , \quad r, r' = L, R \ .
\eea
In the above we have used the short-hand notation $c_L = \sin\theta_{\rm eff}^{(W)}$ and $c_R = \cos\theta_{\rm eff}^{(W)}$.  Then the Lorentz-invariant expression is 
\bea
&&\sum_{r,r'=L, R} W_{\mu\rho} T_{rr'}^{\mu\rho}  = 8 c_L^2 \,(p_l \cdot \tilde{p}^-_{\tilde\chi^+ } ) (p_\nu \cdot p_{\tilde\chi^0}) +8 c_R^2\, (p_\nu \cdot \tilde{p}^+_{\tilde\chi^+} ) (p_l \cdot p_{\tilde\chi^0}) \nonumber \\
&& \qquad -4 c_L c_R m_{\tilde\chi^0} \left[ p_W^2 m_{\tilde\chi^+}  -2  (p_l\cdot s )(p_\nu \cdot p_{\tilde\chi^+}) +2  (p_l\cdot p_{\tilde\chi^+})(s \cdot p_\nu)\right]  \ ,
\eea
where
\be 
\tilde{p}^\pm_{\tilde\chi^+} \equiv p_{\tilde\chi^+}  \pm m_{\tilde\chi^+} s\ . 
\ee
Assuming an on-shell $W$ boson and using the Narrow Width Approximation, we compute the normalized doubly differential spectra in the chargino rest frame
\bea
\label{eq:double}
\frac1{\Gamma} \frac{d\Gamma}{dE_l d\cos\theta_l} &=& 8 m_{\tilde\chi^+}^2 E_l E_\nu (c_L^2+c_R^2) - 4 m_{\tilde\chi^+} m_W^2 \left(  c_L^2 E_l + c_R^2 E_\nu + c_Lc_R m_{\tilde\chi^0}\right) \nonumber \\
&& \quad+ 4 m_{\tilde\chi^+} |\vec{s}| \cos\theta_l \left[ c_L E_l  (2c_L m_{\tilde\chi^+} E_\nu - c_L m_W^2 -2c_R m_{\tilde\chi^0} E_\nu)\phantom{\left(\frac{m_W^2}{2E_l}\right)}\right.\nonumber \\
&&\quad \left. + c_R \left(\frac{m_W^2}{2E_l}-E_\nu\right) \left( 2c_R m_{\tilde\chi^0} E_l - c_R m_W^2 -2c_L m_{\tilde\chi^0} E_l \right) \right] \ ,
\eea
where $\cos\theta_l = \hat{p}_l\cdot \hat{s}$ and
 \be
E_\nu= E_W - E_l= \frac12 \left(m_{\tilde\chi^+} + \frac{m_W^2-m_{\tilde\chi^0}^2}{m_{\tilde\chi^+}}\right) - E_l\ .
 \ee
The doubly spectra agree with the similar expressions for a polarized top decay with anomalous $Wtb$ couplings in  Ref.~\cite{Jezabek:1994zv}, after taking the limit $m_{\tilde\chi^0}\to 0$.  

Integrating over the charged-lepton energy, we arrive at the normalized leptonic angular spectrum
\be
\frac1{\Gamma}\frac{d\Gamma}{d\cos\theta_l} = \frac12 \left(1+\frac{S_2(m_{\tilde\chi^+},m_{\tilde\chi^0}, m_W)}{S_1 (m_{\tilde\chi^+},m_{\tilde\chi^0}, m_W)}\ |\vec{s}| \cos\theta_l \right)\ ,
\ee
where
\bea
&& S_1(m_{\tilde\chi^+},m_{\tilde\chi^0}, m_W) =
 \frac{\lambda^{1/2}(m_{\tilde\chi^+}^2, m_{\tilde\chi^0}^2, m_W^2)}{2m_{\tilde\chi^+}} \left\{ -2 m_{\tilde\chi^+}m_{\tilde\chi^0} m_W^2\, \sin2\theta_{\rm eff}^{(W)}  \right.\nonumber \\
&&\qquad \qquad \left. +\frac13 \left[(m_{\tilde\chi^+}^2-m_{\tilde\chi^0}^2)^2+(m_{\tilde\chi^+}^2+m_{\tilde\chi^0}^2)m_W^2-2m_W^4\right] \right\} \ , \\
&&S_2(m_{\tilde\chi^+},m_{\tilde\chi^0}, m_W) 
= \frac{\lambda^{1/2}(m_{\tilde\chi^+}^2, m_{\tilde\chi^0}^2, m_W^2)}{2m_{\tilde\chi^+}}\left\{  4m_{\tilde\chi^+}^2m_W^2\,  \cos^2 \theta_{\rm eff}^{(W)}  -2 m_{\tilde\chi^+}m_{\tilde\chi^0} m_W^2\, \sin2\theta_{\rm eff}^{(W)}  \phantom{\frac13} \right.\nonumber \\
&& \qquad \qquad \left.-\frac13 \left[(m_{\tilde\chi^+}^2-m_{\tilde\chi^0}^2)^2+(m_{\tilde\chi^+}^2+m_{\tilde\chi^0}^2)m_W^2-2m_W^4\right]  \cos 2\theta_{\rm eff}^{(W)} \right\}\nonumber \\
&&\qquad\qquad-2m_{\tilde\chi^+} \cos^2 \theta_{\rm eff}^{(W)}  m_W^4 \log\left[\frac{m_{\tilde\chi^+}^2+m_W^2-m_{\tilde\chi^0}^2+ \lambda^{1/2}(m_{\tilde\chi^+}^2, m_{\tilde\chi^0}^2, m_W^2)}{ m_{\tilde\chi^+}^2+m_W^2-m_{\tilde\chi^0}^2- \lambda^{1/2}(m_{\tilde\chi^+}^2, m_{\tilde\chi^0}^2, m_W^2)}\right]\ .
\eea
As a check, the case of  top decays can be obtained from the limit $\cos\theta_{\rm eff}^{(W)}=0$, leading to the $S_2/S_1=1$, which is the well-known result that the charged lepton in the top decays has the maximal spin analyzing power. These results also agree with similar computations for a polarized top decays with anomalous $Wtb$ couplings in Ref.~\cite{AguilarSaavedra:2006fy}, which included the finite $m_b$ effect.

\subsection{Spin Analyzing Power}
\label{sect:sub3}

Using results from the previous two subsections, we arrive at the angular distributions of the charged lepton coming out of the stop decays in  the rest frames of the top and the chargino:
\be
\label{eq:leptonspectra}
\frac1{\Gamma}\frac{d\Gamma}{d\cos\theta_l} = \frac12 \left(1+ {\cal P}_l \cos\theta_l \right)
\ee
where 
\bea
\label{eq:polt}
{\cal P}_l^{(t)} &=& \frac{|\vec{p}_{\tilde \chi^0}| \cos 2\theta_{\rm eff}^{(t)} }{E_{\tilde \chi^0}+m_{\tilde \chi^0} \sin 2\theta_{\rm eff}^{(t)}}  \ , \\
{\cal P}_l^{(\chi)} &=&  \frac{|\vec{p}_{b}| \cos 2\theta_{\rm eff}^{(\chi)} }{E_b-m_{b} \sin 2\theta_{\rm eff}^{(\chi)}} \times \frac{S_2(m_{\tilde\chi^+},m_{\tilde\chi^0}, m_W)}{S_1 (m_{\tilde\chi^+},m_{\tilde\chi^0}, m_W)}
\label{eq:chispectra} \\
&\rightarrow&   \cos 2\theta_{\rm eff}^{(\chi)} \times \frac{S_2(m_{\tilde\chi^+},m_{\tilde\chi^0}, m_W)}{S_1 (m_{\tilde\chi^+},m_{\tilde\chi^0}, m_W)} \quad {\rm for} \quad m_b\to 0 \ .
\label{eq:zeromb}
\eea
In the above the energy and momentum of the $b$ quark in the chargino channel and of the neutralino in the top channel are both fixed, as shown in Eqs.~(\ref{eq:ebrest}) and (\ref{eq:pbrest}). Moreover, $\theta_l$ is defined as the angle between the charged lepton and the $b$ quark (neutralino) in the rest frames of the chargino (top).\footnote{We note that our definition of $\theta_l$ differs by $\pi$ from some previous studies on effects of top polarizations in stop decays.}

\begin{figure}[t]
\includegraphics[scale=0.7, angle=0]{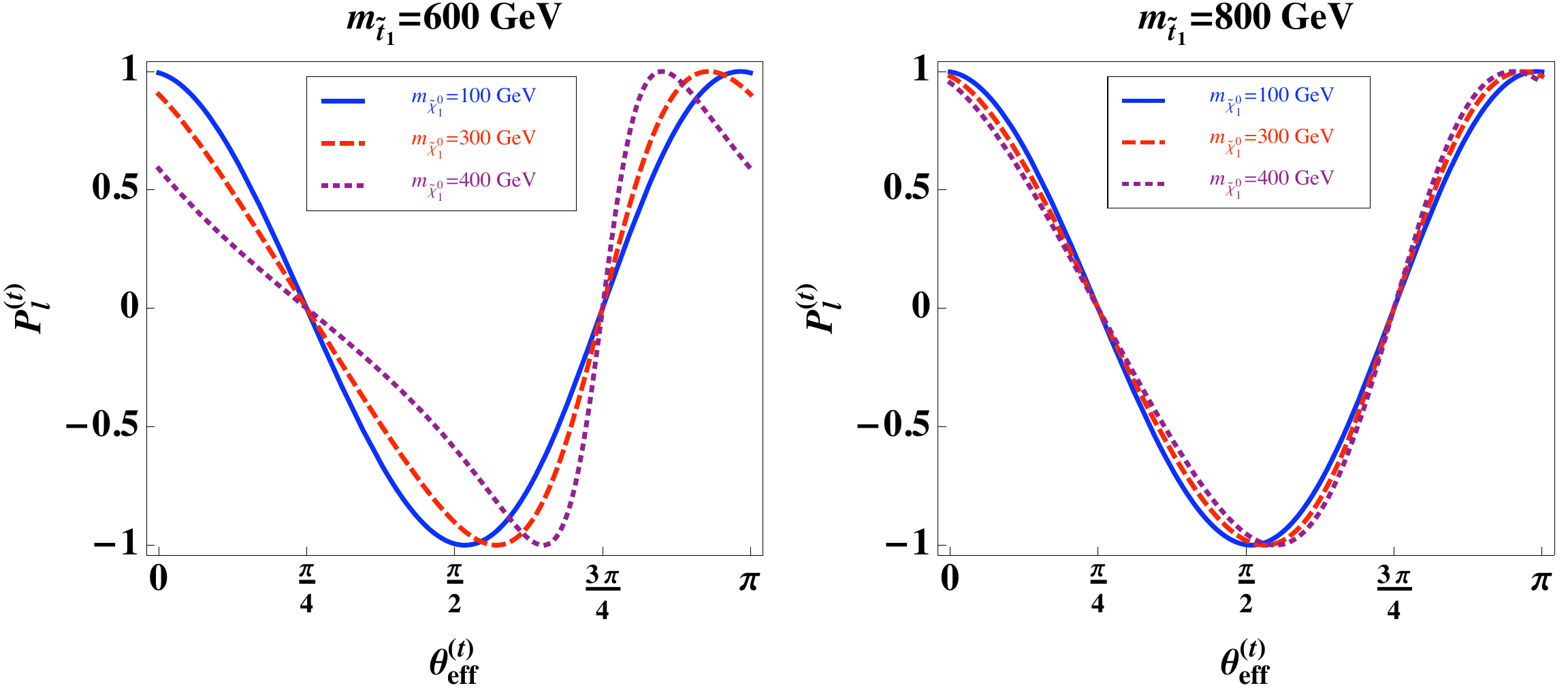}  
\caption{\label{fig1}{\em Spin analyzing power of the charged lepton from the decay $\tilde{t}_1\to t\tilde{\chi}^0_1 \to (l^+\nu b)\tilde{\chi}^0_1$. $P_l^{(t)}$ becomes sensitive to only  the effective mixing angle $\theta_{\rm eff}^{(t)}$ when $m_{\tilde t_1} \gg m_{\tilde \chi^0_1}$.
}}
\end{figure}

In Fig.~\ref{fig1} we plot the spin analyzing power of the charged lepton in the top channel ${\cal P}^{(t)}_l$ as a function of the effective mixing angle $\theta_{\rm eff}^{(t)}$, for two different stop masses $m_{\tilde t_1}=600$ and 800 GeV and three different neutralino masses $m_{\tilde\chi^0_1}=100$, 300, and 500 GeV. We see that, in the limit $m_{\tilde t_1}\gg m_{\tilde \chi_1^0}$, ${\cal P}^{(t)}_l$ only depends on the effective mixing angle and is not sensitive to masses of the particles involved. In this limit  Eq.~(\ref{eq:polt}) becomes:
\be
{\cal P}_l^{(t)} \rightarrow \cos 2\theta_{\rm eff}^{(t)} \ ,
\ee
which is due to the fact that the neutralino becomes highly boosted and, as a result, behaves like a massless particle: $|\vec{p}_{\tilde\chi^0}| \approx E_{\tilde\chi^0} \gg m_{\tilde\chi^0}$.

\begin{figure}[t]
\includegraphics[scale=0.7, angle=0]{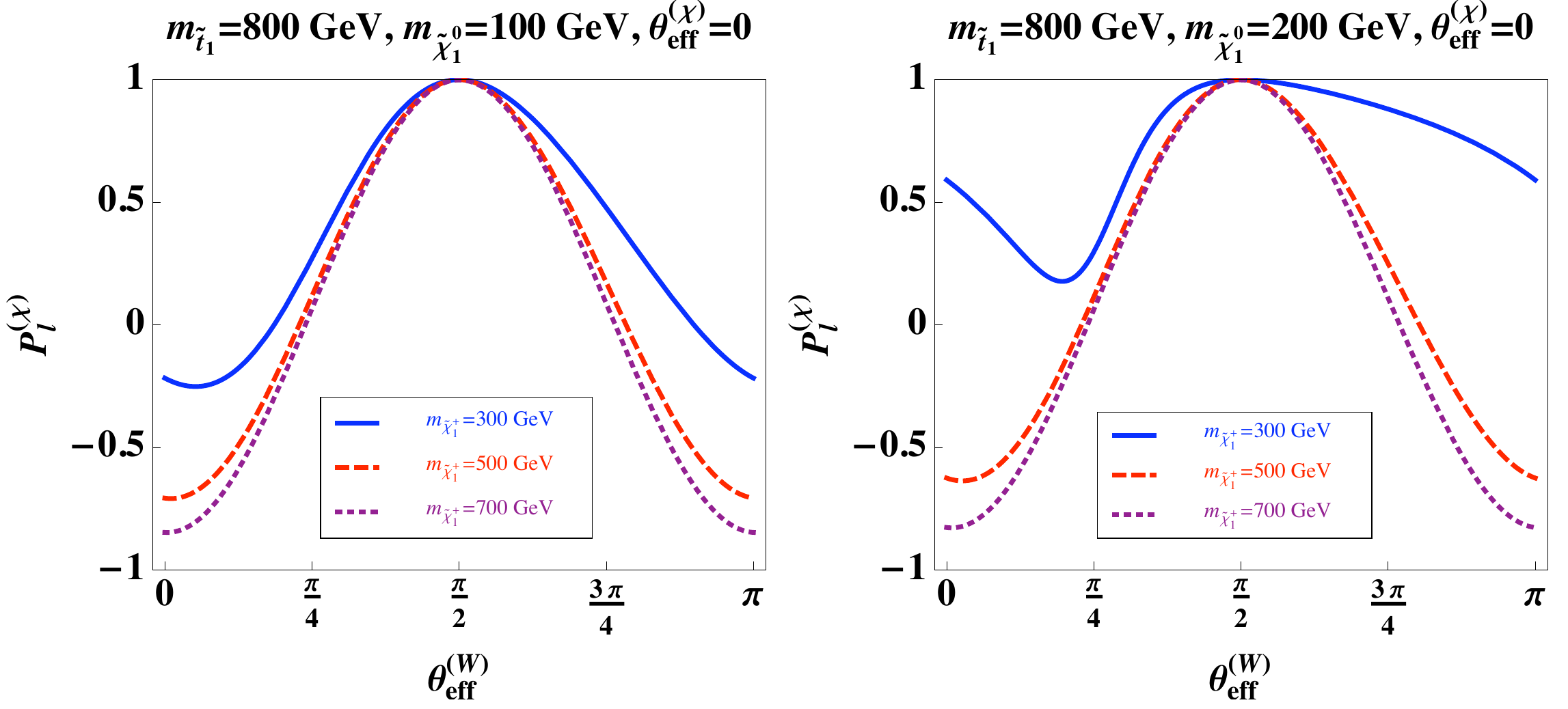}  
\caption{\label{fig2}{\em Spin analyzing power of the charged lepton from the decay $\tilde{t}_1\to b\tilde{\chi}^+_1 \to b(l^+\nu \tilde{\chi}^0_1)$.
}}
\end{figure}

In Fig.~\ref{fig2} we plot the spin analyzing power of the charged lepton in the chargino channel ${\cal P}^{(\chi)}_l$ as a function of the effective mixing angle $\theta_{\rm eff}^{(W)}$. Effects of finite $b$ quark mass is not enhanced by $\tan\beta$ in the two-body decay kinematics of $\tilde{t}_1 \to b\tilde{\chi}^+_1$ and thus can be safely neglected.\footnote{The $\tan\beta$ enhanced $b$ quark mass effect is fully captured in $\theta_{\rm eff}^{(\chi)}$, as discussed in the paragraph containing Eq.~(\ref{eq:ybtan}).} Then two observations can be made from Eq.~(\ref{eq:zeromb}): 1) ${\cal P}_l^{(\chi)}$ becomes independent of the stop mass $m_{\tilde t_1}$, and 2) ${\cal P}_l^{(\chi)} \leftrightarrow -{\cal P}_l^{(\chi)}$ for $\theta_{\rm eff}^{(\chi)} \leftrightarrow \pi/2 - \theta_{\rm eff}^{(\chi)}$. Therefore we only showed plots for $\theta_{\rm eff}^{(\chi)}=0$, which corresponds to a fully polarized chargino in the left-handed helicity eigenstate in the rest frame of the chargino. As discussed previously, this is always the case except in the large $\tan\beta \sim m_t/m_b$. The limit of a right-handed chargino can be obtained by flipping the sign of the spin analyzing power.  From Fig.~\ref{fig2} we see that, in general, the spin analyzing power of the charged lepton is quite sensitive to both the chargino and the neutralino masses,  except when $m_{\tilde\chi_1^+} \gg m_{\tilde\chi_1^0}$.


\section{Lab Frame Observables  at the LHC}
\label{sect:lhc}


In this section we use Monte Carlo simulations to study impacts of chargino and top polarizations in direct stop searches at the LHC, as well as kinematic variables in the laboratory (Lab) frame which would allow for discrimination between the top channel and the chargino channel. We will use the following benchmark scenarios:
\begin{itemize}

\item TopL1: $m_{\tilde t_1}$ = 600 GeV, $m_{\tilde \chi_1^0}$ = 100 GeV, and $\theta_{\rm eff}^{(t)}=\pi/2$. \\
         TopL2: $m_{\tilde t_1}$ = 800 GeV, $m_{\tilde \chi_1^0}$ = 100 GeV, and $\theta_{\rm eff}^{(t)}=\pi/2$.

\item TopR1: $m_{\tilde t_1}$ = 600 GeV, $m_{\tilde \chi_1^0}$ = 100 GeV, and $\theta_{\rm eff}^{(t)}=0$. \\
          TopR2: $m_{\tilde t_1}$ = 800 GeV, $m_{\tilde \chi_1^0}$ = 100 GeV, and $\theta_{\rm eff}^{(t)}=0$.
          
\item ChaL1: $m_{\tilde t_1}$ = 600 GeV, $m_{\tilde \chi_1^+}$ = 300 GeV, $m_{\tilde \chi_1^0}$ = 100 GeV, $\theta_{\rm eff}^{(\chi)}=0$, and $\theta_{\rm eff}^{(W)}=\pi/2$.\\
          ChaL2: $m_{\tilde t_1}$ = 800 GeV, $m_{\tilde \chi_1^+}$ = 700 GeV, $m_{\tilde \chi_1^0}$ = 100 GeV, $\theta_{\rm eff}^{(\chi)}=0$, and $\theta_{\rm eff}^{(W)}=\pi/2$.

\item ChaR1: $m_{\tilde t_1}$ = 600 GeV, $m_{\tilde \chi_1^+}$ = 300 GeV, $m_{\tilde \chi_1^0}$ = 100 GeV, $\theta_{\rm eff}^{(\chi)}=0$, and $\theta_{\rm eff}^{(W)}=0$.\\
          ChaR2: $m_{\tilde t_1}$ = 800 GeV, $m_{\tilde \chi_1^+}$ = 700 GeV, $m_{\tilde \chi_1^0}$ = 100 GeV, $\theta_{\rm eff}^{(\chi)}=0$, and $\theta_{\rm eff}^{(W)}=0$.

\end{itemize}
We generate 20,000 parton level events using {\tt Madgraph 5} \cite{Alwall:2011uj} for each of the  benchmarks. The simulations are validated by comparing the angular spectra of the charged lepton with those in Eqs.~(\ref{eq:leptonspectra}) - (\ref{eq:chispectra}).  In particular, we focus on events where $W^+$ decays leptonically and $W^-$ decays hadronically so that the final states are
\be
pp \to \tilde{t}_1 \tilde{t}_1^* \to 2j +  2b + l^+ + \slashchar{E}_T
\ee
for both the top and the chargino channels. The cross-sections for stop pair production at 8 (13) TeV LHC range from 0.025 (0.17) pb for $m_{{\tilde t}_1}=600$ GeV to 0.0029 (0.028)  pb for $m_{{\tilde t}_1}=800$ GeV \cite{Kramer:2012bx, susy}. The dominant background to the signal process comes from SM top quark pair productions, when the both $W$ bosons decay leptonically with one of the leptons not identified, or one $W$ decays hadronically and the other leptonically. Two key observables to separate the dominant background from the signal are the missing transverse energy and the transverse mass \cite{Han:2008gy}. The missing transverse energy in the signal is all coming from the neutrinos, which are almost massless, while that in the signal arises from both neutrinos and the massive neutralinos. In addition, because the charged lepton in the background comes from the $W$ boson, its transverse mass is expected to exhibit a Jacobian peak at $m_W$, while the $m_T$ distribution from the signal can extend far above the $W$ mass \cite{Han:2008gy}. Therefore in our simulation we adopt an aggressive cut on both the missing transverse energy and the transverse mass to reduce the background, as summarized in Table \ref{tab:spin}. These cuts are similar to those adopted in the cut-based analyses from both ATLAS and CMS \cite{ATLAS-CONF-2012-166, CMS-PAS-SUS-12-023}. However, since our main interest is to study effects of polarizations on the acceptance rates in experimental searches for direct stop production at the LHC, we would focus on generating signal events only in the Monte Carlo study. A full simulation including the background is obviously beyond the scope of the present work.

\begin{figure}[t]
\includegraphics[scale=0.64, angle=0]{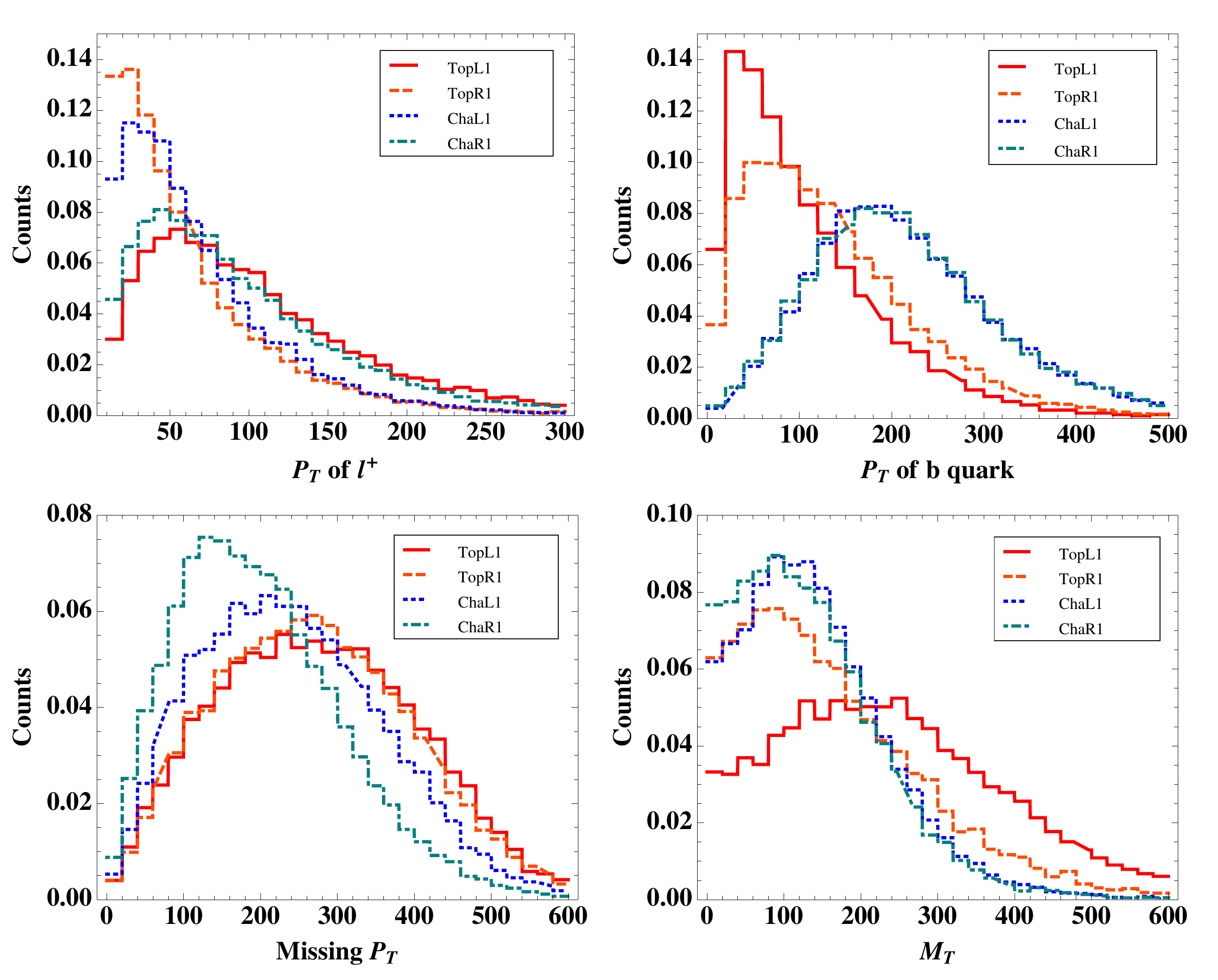}  
\caption{\label{fig3}{\em Kinematic distributions  of $\tilde{t}_1\to bl^+\nu \tilde{\chi}^0_1$ at the parton level, without any selections.
}}
\end{figure}

In Fig.~\ref{fig3} we plot, for the purpose of illustration, spectra of four kinematic variables used in current searches for the benchmark scenarios with $m_{\tilde t_1}=600$ GeV. We then impose kinematic cuts on the four kinematic variables, as  summarized in Table \ref{tab:cuts}.  The cut flows as well as the spin-analyzing powers for the charged leptons for each of the benchmark scenarios are listed in Table \ref{tab:spin}, from which we see that the polarization of the top quark could potentially have a significant impact on the selection efficiencies in current searches and the resulting limits on stop masses, while the impact of the chargino polarization seems less significant. In particular, Cut 1 in Table \ref{tab:cuts} has the strongest dependence on the top polarization. This dependence is mostly due to the $p_T$ spectrum of the charged lepton, as can be seen from Fig.~\ref{fig3}.

\begin{table}
\begin{center}
  \begin{tabular}{l| p{7cm} l} \hline \hline 
 Cut 1  & $p_T \ge 30$ GeV and $|\eta| \le 2.4 $ for both the charged lepton and the $b$ quark     \\ \hline 
   Cut 2  & $\slashchar{E}_T \ge 150$ GeV and $M_T \ge 120$ GeV     \\ 
      \hline \hline
  \end{tabular}
\end{center}
\caption{\em Parton level cuts to study impacts of polarizations on kinematic variables.  \label{tab:cuts}}
\end{table}

\begin{table}
\begin{center}
\begin{tabular}{|c|c|c|c|c|c|c|c|c|}
\hline
    $\sqrt{s}=8$  TeV   & TopL1 & TopR1  & TopL2 & TopR2 & ChaL1 & ChaR1 & ChaL2 & ChaR2   \\ \hline
 Events & 20,000 & 20,000 & 20,000 & 20,000 & 20,000 & 20,000 & 20,000 & 20,000 \\ \hline 
 Cut 1     &  15,508 & 12,316 & 16,313   &  13,118  & 14,765 & 16,996 & 17,855 & 18,100  \\ \hline
  Cut 2     &  11,226 &  8,117 & 13,409   &  10,092  & 7,586 & 6,408 & 13,922 & 13,719 
       \\ \hline    \hline
 \quad $ {\cal P}_l \quad$    & $\ \ -0.99\ \ $ & $\ \  +0.99\ \ $  & $\ \ -1.00\ \ $ & $\ \ +1.00\ \ $ & $\ \ +1.00\ \ $ & $\ \ -0.22\ \ $ & $\ \ +1.00\ \ $ & $\ \ -0.85\ \ $   \\ \hline  
\end{tabular}
\end{center}
\caption{\em Cut flows and  spin analyzing powers of the charged lepton for the benchmark scenarios.}\label{tab:spin}
\end{table}

It should be noted that, in the current study, we do not include effects of the hadronization of the quark in the final states as well as any detector resolutions on the energies, because the main purpose here is to understand how polarizations could affect the kinematic variables used in experimental  selection cuts without introducing additional biases from hadronization and detector resolution. In addition, we do not impose any cuts on the decay products of $\tilde{t}_1^*$ since there is no correlation between decay products of $\tilde{t}_1$ and $\tilde{t}_1^*$. Our results suggest that fully realistic simulations including effects of polarizations, especially in the top channel, is warranted and deserve further scrutiny.

\begin{figure}[t]
\includegraphics[scale=0.7, angle=0]{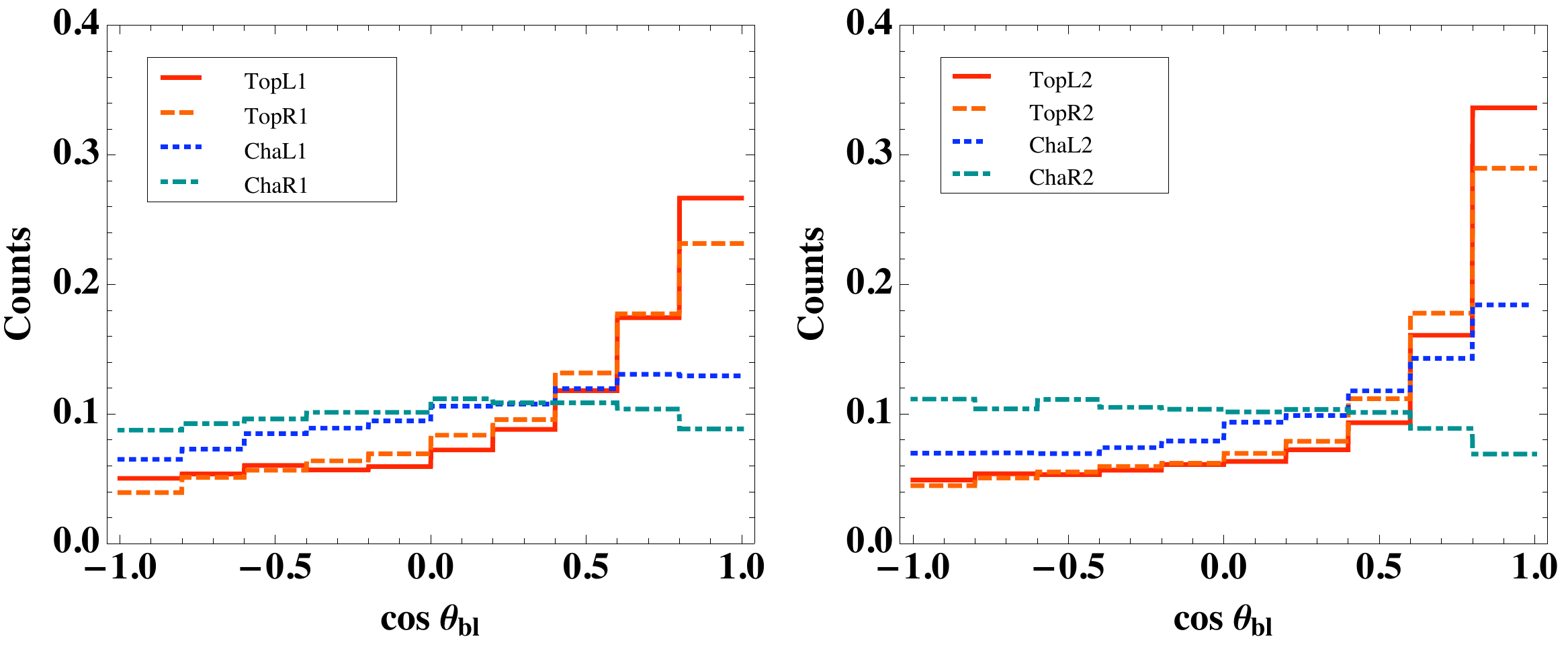}  
\caption{\label{fig4}{\em Cosine of $\theta_{bl}$, the opening angle between the charged lepton and the $b$ quark in the Lab frame. Results shown here are after the selection cuts in Table \ref{tab:cuts} and take into account the combinatorial factor of not being able to distinguish a $b$ quark from a $\bar{b}$ quark.
}}
\end{figure}

\begin{figure}[t]
\includegraphics[scale=0.64, angle=0]{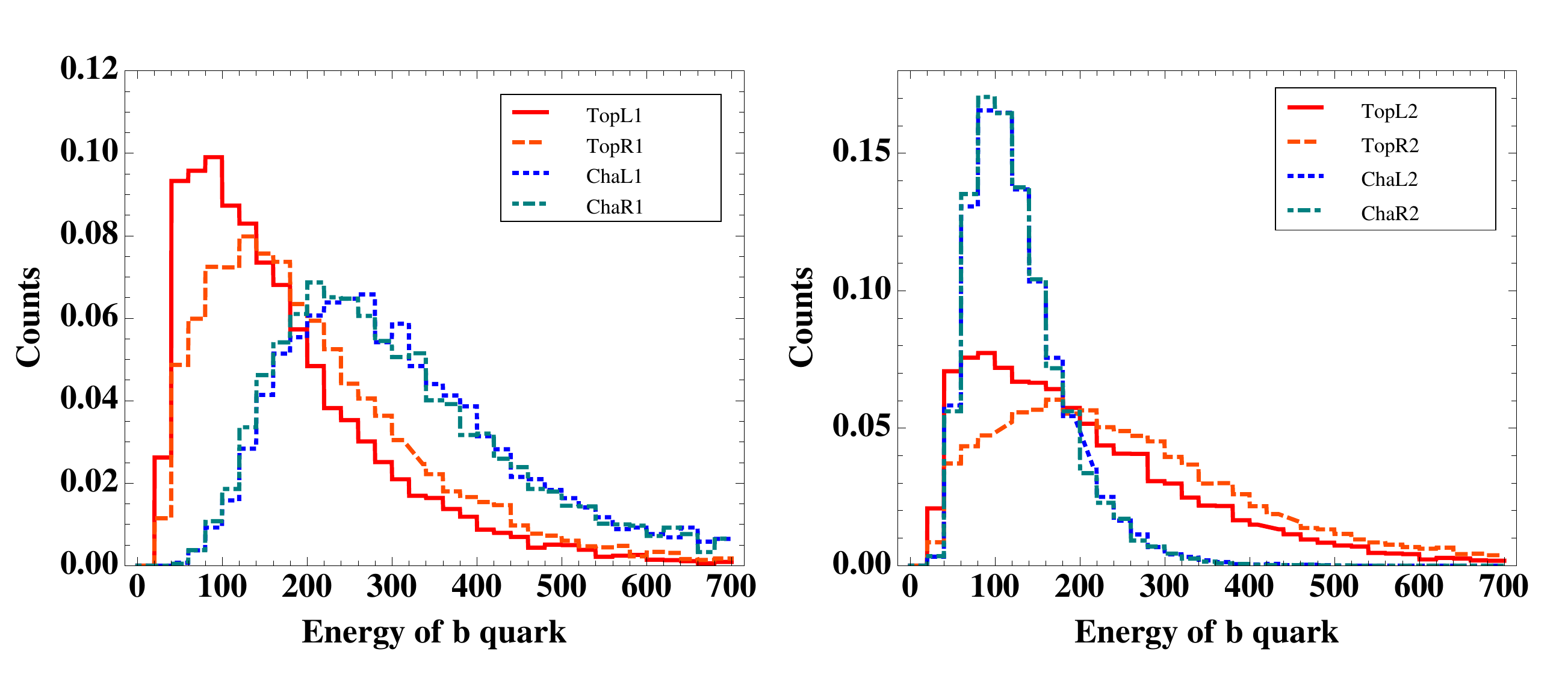}  
\caption{\label{fig5}{\em Energy distributions of the $b$ quark in the Lab frame after the selection cuts in Table \ref{tab:cuts}. Notice that there is no combinatorial factor in this case since the energy distributions are identical for both $b$ and $\bar{b}$ quarks.
}}
\end{figure}

Next we consider two Lab frame observables which could differentiate between the chargino channel from the top channel in stop decays. The two variables we consider are 1) $\theta_{bl}$ the opening angle between the charged lepton and the $b$ quark and 2) $E_b$ the energy of the $b$ quark in the Lab frame. The results, after the selection cuts in Table ~\ref{tab:cuts}, are shown in Figs.~\ref{fig4} and \ref{fig5}. In particular, since experimentally it is very difficult to measure the charge of the $b$ quark on an event-by-event basis, the plot for $\cos \theta_{bl}$ takes into account the combinatorial factor of not being able to distinguish the $b$ quark from the $\bar{b}$ quark.

In Fig.~\ref{fig4} we see that $\cos\theta_{bl}$ from the top channel $\tilde{t}_1\to t\tilde{\chi}^0 \to (W^+b)\tilde{\chi}^0$ is in general much smaller than from the chargino channel $\tilde{t}_1\to b\tilde{\chi}^+ \to b(W^+ \tilde{\chi}^0)$. The physics is simple to understand: the charged lepton and the $b$ quark both come from the top decays in the top channel, and tend to move in the same direction because of the boost of the top quark. This is to be contrasted with the chargino channel where the $b$ quark originates from the 2-body decays of the stop. Moreover, in the rest frame of the chargino the direction of the $b$ quark is the natural axis of polarization for the chargino. We see from comparing the two benchmarks of ChaR1 and ChaR2 that the opening angle in the Lab frame does retain some sensitivity to the polarization of the chargino. It is also worth noting that the major irreducible background for the stop decays is the standard model $t\bar{t}$ production, in which case the $b$ quark and the charged lepton are also from the top decays and tend to move in the same direction, similar to the case of the top channel of stop decays.

Given that $\cos\theta_{bl}$ retains sensitivity to the polarization of the chargino, in the event of discovery one could define the forward-backward asymmetry:
\be
{A}_{FB}^{(bl)} = \frac{\left(\int_0^{1}- \int_{-1}^0 \right) d\cos\theta_{bl} \frac{d\sigma}{d\cos\theta_{bl}}}{ \left(\int_{-1}^0 +\int_0^{1}\right) d\cos\theta_{bl} \frac{d\sigma}{d\cos\theta_{bl}}}
\ee
to measure the polarization, and hence the mixing angles defined in Eqs.~(\ref{eq:bchareff}) -- (\ref{eq:tanWeff}).

Following the same observation that the $b$ quark in the top channel originates from the top quark while that in the chargino channel comes from the stop decays, the kinematics of the $b$ quark could in principle help distinguish between those $b$'s that are from the stop decays versus those from the top decays. In Fig.~\ref{fig5} we plot the energy distributions of the $b$ quark in the Lab frame for the benchmarks we consider. Notice that, unlike the angular spectrum in $\theta_{bl}$, there is no combinatorial factor due to not being able to tell the $b$ from the $\bar{b}$, since both of them should have identical energy spectra. From Fig.~\ref{fig5} we see the simulations bear out the intuition that the $b$ quark kinematics behave differently in the top and the chargino channels. In fact,  the energy of the $b$ quark from the top decays is peaked not far from the fixed energy in the rest frame of the top quark, which is
\be
\label{eq:btop}
\frac{m_t^2-m_W^2-m_b^2}{2m_t} \approx 70 \ \ {\rm GeV}\ .
\ee
It was shown in Ref.~\cite{Agashe:2012bn} that for {\em unpolarized} top quark decays, the peak in the energy spectrum of the $b$ quark is invariant under any boost and exactly as in Eq.~(\ref{eq:btop}). In our case, the top quark is not completely unpolarized and one does see some dependence on the top polarizations in Fig.~\ref{fig5}. Nevertheless, it is worth pointing out that the top quark produced in the standard model $t\bar{t}$ production is unpolarized and the energy spectrum of the $b$ quark from this irreducible background would have a peak very close to the value predicted by Eq.~(\ref{eq:btop}).  For the $b$ quark from the chargino decays, its kinematics is largely determined by the stop decays and the peak in the energy distribution is again not far from the fixed energy in the rest frame of the stops given by
\be
\label{eq:bcha}
\frac{m_{\tilde t_1}^2-m_{\tilde \chi^+}^2-m_b^2}{2m_{\tilde{t}_1}} \ .
\ee
Therefore in ChaL1/ChaR1 benchmarks the peak is at approximately 225 GeV while it is at around 95 GeV for ChaL2/ChaR2, which seem to give pretty accurate locations of the maximum $b$ energy in Fig.~\ref{fig5}. Moreover, the shift in the peaks due to the chargino polarization is much less than the corresponding shift due to the top polarization.


\section{Conclusions}


In this work we considered polarizations issues in stop decays in both the top and the chargino channels. Energy and angular spectra of the charged lepton were presented and the possibility of the charged lepton as the chargino spin-analyzer was studied.

We also performed parton-level Monte Carlo simulations to study impacts of polarizations on kinematic variables in the laboratory frame at the LHC. We found that the selection efficiencies in the top channel could be affected significantly by the polarization, while the corresponding effect in the chargino channel is less significant. In addition, we proposed two variables in the laboratory frame, $\cos\theta_{bl}$ and $E_b$, to optimize the searches in the chargino versus the top channels. Our results suggest  simulations including full detect effects and relevant backgrounds are warranted and should be undertaken.

 \begin{acknowledgements}

I am grateful to Claudio Campagnari for motivating this study and helpful discussions. I also benefitted from discussions with Kaustubh Agashe, Juan Alcaraz and Xerxes Tata.  In addition, I thank Kaustubh Agashe for pointing out Ref.~\cite{Agashe:2012bn}.  Hospitality at the Mainz Institute for Theoretical Physics during the workshop "The first three years of the LHC" is acknowledged while this manuscript was being completed.
This work is supported in part by DOE under Contract No. DE-AC02-06CH11357 (ANL) and No. DE-FG02-91ER40684 (Northwestern), and by the Simons Foundation under award No. 230683. Work at KITP is supported by the National Science Foundation under Grant No. PHY11-25915. 

 \end{acknowledgements}


\begin{thebibliography}{nn}


\bibitem{ATLAS-CONF-2012-166} 
  [ATLAS Collaboration],
  ATLAS-CONF-2012-166;
   [ATLAS Collaboration],
  ATLAS-CONF-2013-037.



\bibitem{CMS-PAS-SUS-12-023} 
  [CMS Collaboration],
  CMS-PAS-SUS-12-023.


\bibitem{ATLAS-CONF-2013-061} 
  [ATLAS Collaboration],
  ATLAS-CONF-2013-061;

\bibitem{CMS-PAS-SUS-13-011} 
 S.~Chatrchyan {\it et al.}  [CMS Collaboration],
  arXiv:1308.1586 [hep-ex].

\bibitem{Claudio}
Claudio Campagnari, private communications.


\bibitem{Kitano:2002ss} 
  R.~Kitano, T.~Moroi and S.~-f.~Su,
  JHEP {\bf 0212}, 011 (2002)
  [hep-ph/0208149].


\bibitem{Shelton:2008nq} 
  J.~Shelton,
  Phys.\ Rev.\ D {\bf 79}, 014032 (2009)
  [arXiv:0811.0569 [hep-ph]].

\bibitem{Perelstein:2008zt} 
  M.~Perelstein and A.~Weiler,
  JHEP {\bf 0903}, 141 (2009)
  [arXiv:0811.1024 [hep-ph]].

\bibitem{Berger:2012an} 
  E.~L.~Berger, Q.~-H.~Cao, J.~-H.~Yu and H.~Zhang,
  Phys.\ Rev.\ Lett.\  {\bf 109}, 152004 (2012)
  [arXiv:1207.1101 [hep-ph]].


\bibitem{Kilic:2012kw} 
  C.~Kilic and B.~Tweedie,
  JHEP {\bf 1304}, 110 (2013)
  [arXiv:1211.6106 [hep-ph]].

  
\bibitem{Belanger:2012tm} 
  G.~Belanger, R.~M.~Godbole, L.~Hartgring, I.~Niessen and ,
   JHEP {\bf 1305}, 167 (2013)
  [arXiv:1212.3526].


\bibitem{Bai:2013ema} 
  Y.~Bai, H.~-C.~Cheng, J.~Gallicchio and J.~Gu,
JHEP {\bf 1308}, 085 (2013)
  [arXiv:1304.3148 [hep-ph]].
  
\bibitem{Belanger:2013gha} 
  G.~Belanger, R.~M.~Godbole, S.~Kraml and S.~Kulkarni,
  arXiv:1304.2987 [hep-ph].


\bibitem{Jezabek:1988ja} 
  M.~Jezabek and J.~H.~Kuhn,
  Nucl.\ Phys.\ B {\bf 320}, 20 (1989);
  A.~Czarnecki and M.~Jezabek,
  Nucl.\ Phys.\ B {\bf 427}, 3 (1994)
  [hep-ph/9402326].
  
\bibitem{Czarnecki:1990pe} 
  A.~Czarnecki, M.~Jezabek and J.~H.~Kuhn,
  Nucl.\ Phys.\ B {\bf 351}, 70 (1991).
  
  
\bibitem{Jezabek:1994zv} 
  M.~Jezabek and J.~H.~Kuhn,
  Phys.\ Lett.\ B {\bf 329}, 317 (1994)
  [hep-ph/9403366].


\bibitem{Djouadi:2005gj}
See, for example, A.~Djouadi,
arXiv:hep-ph/0503173.

\bibitem{Gunion:1984yn} 
  J.~F.~Gunion and H.~E.~Haber,
  Nucl.\ Phys.\ B {\bf 272}, 1 (1986)
  [Erratum-ibid.\ B {\bf 402}, 567 (1993)].



  
\bibitem{Nojiri:1994it} 
  M.~M.~Nojiri,
  Phys.\ Rev.\ D {\bf 51}, 6281 (1995)
  [hep-ph/9412374].


\bibitem{Juan}
Juan Alcaraz, private communications.

\bibitem{AguilarSaavedra:2006fy} 
  J.~A.~Aguilar-Saavedra, J.~Carvalho, N.~F.~Castro, F.~Veloso and A.~Onofre,
  Eur.\ Phys.\ J.\ C {\bf 50}, 519 (2007)
  [hep-ph/0605190].

\bibitem{Alwall:2011uj} 
  J.~Alwall, M.~Herquet, F.~Maltoni, O.~Mattelaer and T.~Stelzer,
  JHEP {\bf 1106}, 128 (2011)
  [arXiv:1106.0522 [hep-ph]].
  
  
\bibitem{Kramer:2012bx} 
  M.~Kramer, A.~Kulesza, R.~van der Leeuw, M.~Mangano, S.~Padhi, T.~Plehn and X.~Portell,
  arXiv:1206.2892 [hep-ph].

\bibitem{susy}
LHC SUSY Cross-section Working Group,\\
{\tt https://twiki.cern.ch/twiki/bin/view/LHCPhysics/SUSYCrossSections}

\bibitem{Han:2008gy} 
  T.~Han, R.~Mahbubani, D.~G.~E.~Walker and L.~-T.~Wang,
  JHEP {\bf 0905}, 117 (2009)
  [arXiv:0803.3820 [hep-ph]].

\bibitem{Agashe:2012bn} 
  K.~Agashe, R.~Franceschini and D.~Kim,
 Phys.\ Rev.\ D {\bf 88}, 057701 (2013)
  [arXiv:1209.0772 [hep-ph]].

\end{thebibliography}
\end{document}